\documentclass[pra,aps,twocolumn,showpacs,superscriptaddress]{revtex4}
\usepackage{color} %used for font color    
\usepackage{bm}
\usepackage{times}
\usepackage{amssymb} %maths            
\usepackage{amsmath} %maths                      
\usepackage{ragged2e}
\usepackage{graphicx}
\usepackage{epstopdf}
\usepackage[english]{babel}
\usepackage{mathrsfs}
\usepackage{textcomp}
\usepackage{amsthm}
\usepackage{amsfonts}
\usepackage{soul}
\usepackage{braket} 
\usepackage{hyperref}
\usepackage{cleveref}
\newcommand{\be}{\begin{equation}}
\newcommand{\ee}{\end{equation}}
\newcommand{\bea}{\begin{eqnarray}}
\newcommand{\eea}{\end{eqnarray}}
\newcommand{\ba}{\begin{array}}
\newcommand{\ea}{\end{array}}

\def\bs#1{\boldsymbol{#1}}

\begin{document}

\title{Bose-Hubbard physics in synthetic dimensions from interaction Trotterization}
\author{L. Barbiero}
\email{luca.barbiero@ulb.ac.be}
\affiliation{Center for Nonlinear Phenomena and Complex Systems,
Universit\'e Libre de Bruxelles, CP 231, Campus Plaine, B-1050 Brussels, Belgium}
\author{L. Chomaz} 
\affiliation{Institut f\"ur Experimentalphysik, Universit\"at Innsbruck, Innsbruck, Austria.}
\author{S. Nascimbene}
\affiliation{Laboratoire Kastler Brossel, Coll\`ege de France, CNRS, ENS-PSL University, Sorbonne Universit\'e,
11 Place Marcelin Berthelot, 75005 Paris, France}
\author{N. Goldman}
\email{ngoldman@ulb.ac.be}
\affiliation{Center for Nonlinear Phenomena and Complex Systems,
Universit\'e Libre de Bruxelles, CP 231, Campus Plaine, B-1050 Brussels, Belgium}

\begin{abstract}

Activating transitions between a set of atomic internal states has emerged as an elegant scheme by which lattice models can be designed in ultracold atomic gases. In this approach, the internal states can be viewed as fictitious lattice sites defined along a synthetic dimension, hence offering a powerful method by which the spatial dimensionality of the system can be extended. Inter-particle collisions generically lead to infinite-range interactions along the synthetic dimensions, which a priori precludes the design of Bose-Hubbard-type models featuring on-site interactions. In this article, we solve this obstacle by introducing a protocol that realizes strong and tunable ``on-site" interactions along an atomic synthetic dimension. Our scheme is based on pulsing strong intra-spin interactions in a fast and periodic manner, hence realizing the desired ``on-site" interactions in a digital (Trotterized) manner. We explore the viability of this protocol by means of numerical calculations, which we perform on various examples that are relevant to ultracold-atom experiments. This general method, which could be applied to various atomic species by means of fast-response protocols based on  Fano-Feshbach resonances, opens the route for the exploration of strongly-correlated matter in synthetic dimensions.

\end{abstract}


\maketitle

\section{Introduction}

Quantum simulation offers a method by which complex phenomena can be analyzed using well-designed quantum systems~\cite{Georgescu_review}. Various physical platforms have been envisaged in this vaste program, including ultracold atoms in optical lattices, trapped ions, superconducting qubits, photonic devices, quantum dots and point defects in diamond. Many of these versatile systems can be arranged in lattices of various geometries, allowing for particles to move in $D\!=\!1, 2$ or $3$ spatial dimensions; see for instance the reviews~\cite{Bloch_review,Windpassinger_review}.  Beyond these physical lattices, it was suggested that motion can also be activated along a ``synthetic dimension" associated with internal degrees of freedom~\cite{boada,celi,Ozawa_review}. This ingenious feature substantially extends the quantum-simulation toolbox:~not only does it allow for the study of 2D phenomena (such as the quantum Hall effect) starting from a 1D lattice system~\cite{celi,Salerno}, it also offers the appealing possibility of exploring higher-dimensional ($D\!>\!3$) physics in the laboratory~\cite{boada,Price-Goldman-4D,Ozawa-4D,Price-6D}. 

In ultracold atoms, several schemes have been proposed to increase the effective dimensionality of the system. The original proposals of Refs.~\cite{boada,celi} suggested to use internal atomic states (``spins") as fictitious lattice sites along a synthetic dimension, the motion along which can be activated by coupling these states with Raman transitions. This scheme was then generalized to long-lived electronic orbital states coupled through an optical clock transition~\cite{livi,Kolkowitz}, rotational states of ultracold molecules coupled by microwaves~\cite{sundar}, discrete momentum states coupled by Bragg transitions~\cite{gadway}, and harmonic-oscillator states coupled by shaking of the trap~\cite{price,Salerno}. Experimental realizations of synthetic dimensions revealed diverse phenomena, including chiral edge currents~\cite{mancini,stuhl,an}, Anderson topological insulators~\cite{meier}, topological solitons~\cite{meier2} and a flux-dependent mobility edge in disordered chains~\cite{an2}. In parallel, similar techniques were developed in photonics~\cite{Jukic,Luo,Ozawa-4D,Yuan,Segev,Ozawa_review}, and very recently in the solid state~\cite{Price_solid}.
%%%%

\begin{figure}
\includegraphics[scale=0.3]{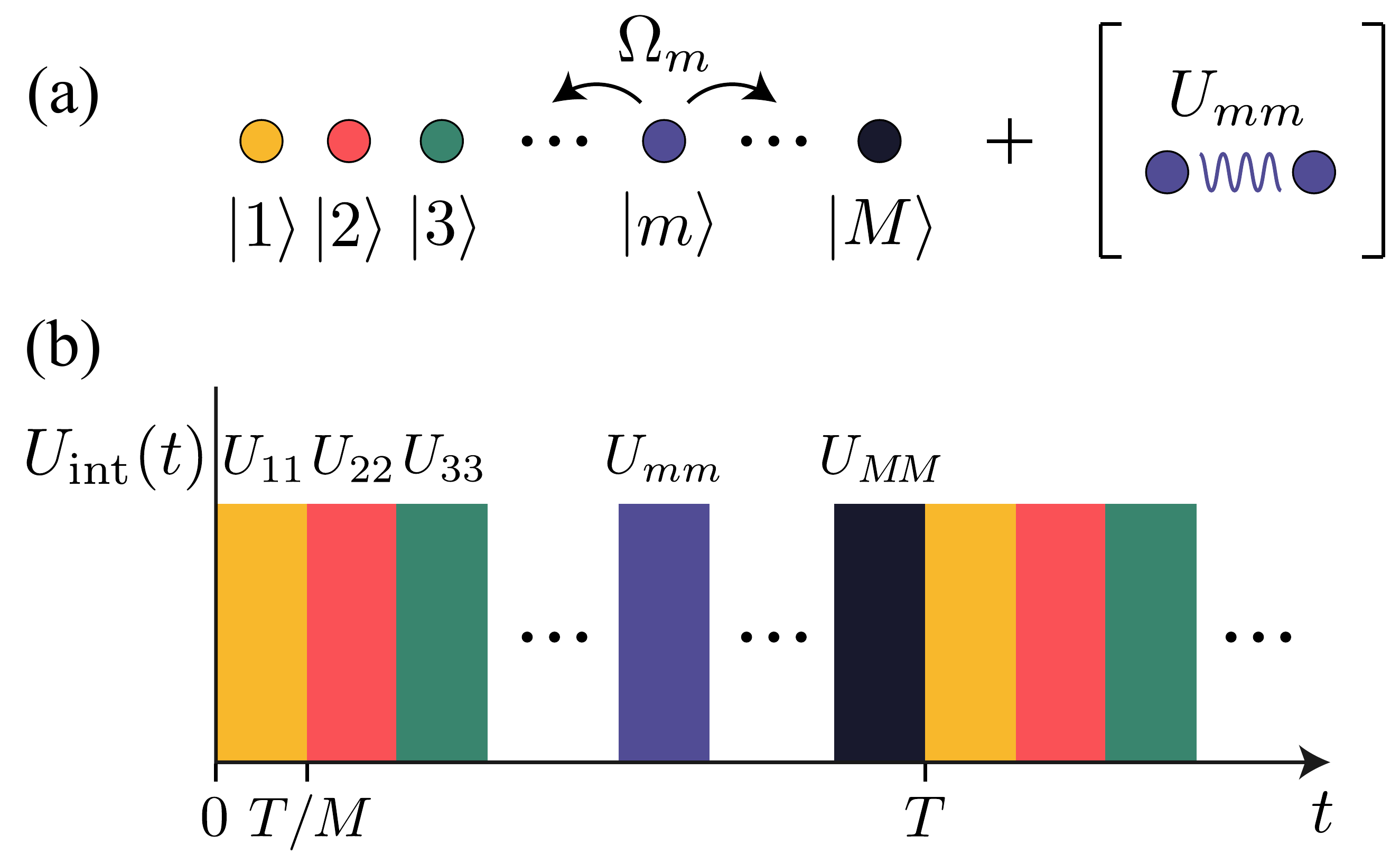}
\caption{(a) A synthetic dimension is spanned by $M$ internal states of an atom. ``Hopping" matrix elements along the synthetic dimension, $\Omega_m$, are controlled by laser coupling, while intra-spin interactions $U_{mm}$ are tuned by Fano-Feshbach resonances. The scheme assumes that inter-spin interactions $U_{m\ne n}$ (infinite-range interactions along the synthetic dimension) can be made small compared to all other energy scales. (b) The Trotterization scheme consists in pulsing intra-spin interactions in a successive and time-periodic manner:~in the first step, one activates $U_{11}$ only, then $U_{22}$ only, etc \dots The drive frequency $\omega\!=\!2 \pi/T$ is assumed to be the largest frequency in the system: $\hbar \omega \gg \Omega_m, U_{mm}$.}
\label{fig1}
\end{figure}

Until now, experiments exploiting atomic synthetic dimensions have been operating in the non-interacting regime, where physical observables can be described in terms of a single-particle band structure. However, recent theoretical works have suggested that intriguing physical phenomena can arise in these settings upon taking (strong) inter-particle interactions into account.  In the case of fermionic species, synthetic dimensions with spin-independent interactions were shown to generate special crystalline states~\cite{barbarino}, exotic edge currents~\cite{yan,cornfeld,barbarino2}, orbital magnetism~\cite{zhang}, and symmetry-protected topological insulating states~\cite{juneman}. In the case of bosonic species, where collisions in the same internal state are also allowed, theorists have predicted supersolid states~\cite{bilitevsky}, vortex-hole duality phenomena~\cite{greschner} and analog Haldane phases~\cite{xu}. 

Importantly, two-body contact interactions generically translate into infinite-range interactions along synthetic dimensions spanned by atomic internal states~\cite{celi}. In fact, for certain atomic species, the scattering length is independent of the atomic internal states involved in the collisions, hence leading to SU($M$)-invariant interactions (where $M$ refers to the number of internal states involved); see Refs.~\cite{Cazalilla,Gorshkov,Pagano}. While infinite-range interactions lead to interesting phenomena, they effectively reduce the dimensionality of the synthetic dimension:~Indeed, in the presence of strong interactions, spin-changing processes can be neglected, and the infinite-range nature of the interactions effectively reduces the synthetic dimension from 1D to 0D. Besides, the effects generated by SU($M$) interactions are generically captured by mean-field approaches. These observations indicate that genuine 2D interacting quantum states, such as fractional quantum Hall (FQH) states~\cite{tsui,laughlin,prange}, cannot be created by combining a real lattice system (with on-site interactions) and a synthetic dimension with infinite-range interactions; the absence of FQH states in such a setting was analyzed numerically in Ref.~\cite{lacki} through the calculation of edge currents. Altogether, this indicates the importance of designing controllable on-site (or finite-range) interactions in synthetic dimensions~\cite{Ozawa_onsite,price}, in view of exploring strongly-interacting topological matter in these settings.

In this work, we propose a realistic scheme realizing effective on-site interactions in a synthetic dimension spanned by atomic internal states. Our method builds on using multiple Fano-Feshbach resonances in atomic spin mixtures~\cite{kaufman_radio-frequency_2009,zhang_independent_2009,chin_feshbach_2010} to select and enhance intra-spin interactions. Specifically, we propose to use this technology to pulse strong intra-spin interactions in a fast and time-periodic manner [Fig.~\ref{fig1}]. As we demonstrate below, this Trotterization scheme effectively generates strong and constant on-site interactions along the synthetic (internal-state) dimension, hence realizing the Bose-Hubbard model in this simple setting. We validate this approach based on several examples that are relevant to ongoing experimental efforts, comparing the dynamics resulting from the full-time-dependent Hamiltonian with that of the effective Hubbard-type model. In particular, we study the impact of the background infinite-range interactions, which will typically be present in experimental implementations. As illustrative examples, we consider the bosonic Mott-superfluid transition and the formation of an effective antiferromagnetic order in a single synthetic dimension, as well as the creation of a chiral Mott-Meissner phase in a synthetic ladder geometry with an effective magnetic flux. The latter setting, which features a combination of effective on-site interactions and artificial magnetic flux, constitutes a promising starting point for the study of FQH physics in synthetic dimensions.

\section{The model and the Trotter protocol}

In this Section, we introduce a general scheme by which on-site interactions can be generated in laser-induced synthetic dimensions, hence realizing the Bose-Hubbard model for cold atoms in the absence of a physical lattice.

As illustrated in Fig.~\ref{fig1}(a), we consider a bosonic atom with $M$ internal states, denoted $\vert m \rangle$ with $m\!=\!(1, \dots, M)$. These (potentially many) internal states, or ``spins", will be viewed as fictitious lattice sites aligned along a synthetic dimension~\cite{Ozawa_review}. The ``hopping" along the synthetic dimension can be activated by coupling these internal states using a proper laser configuration, which then defines the corresponding ``hopping" matrix elements $\Omega_m$. When considering an ensemble of such atoms, the system generically presents a wide variety of interaction processes, which typically depend on the internal states involved in the collisions; in the following, we neglect spin-changing collisions~\cite{note_collisions}. Considering a strongly confined atomic sample, such that its motional degrees of freedom are frozen in the ground state of the confining potential (`single-mode approximation'), we describe such a system by the extended Bose-Hubbard Hamiltonian
\begin{eqnarray}
\hat H=&-&\sum_{m=1}^M (\Omega_{m} \hat b^{\dagger}_{m+1} \hat b_{m} + \text{h.c.})\\
\nonumber
&+&\sum_{m}\frac{U_{mm}}{2} \hat n_{m}(\hat n_{m}-1)+\sum_{m\neq n}\frac{U_{mn}}{2} \hat n_m \hat n_n ,
\label{eq1}
\end{eqnarray}
where $\hat b^{\dagger}_{m}$ (resp.~$\hat n_m$) is the creation (resp.~number) operator of a boson in the internal state $m$, and where the coefficients $U_{mn}$ characterize the various inter-spin ($m\!\ne\!n$) and intra-spin ($m\!=\!n$) interaction strengths. Specifically, these coefficients are given by $U_{mn}\!=\!a_{mn}(4 \pi \hbar^2  /M) \int \text{d}^3 \bs{x} \, w^4(\bs x)$, where $a_{mn}$ is the scattering length associated with spin-conserving collisions in internal states $m$ and $n$, where $M$ is the mass of an atom, and where $w(\bs x)$ denotes the harmonic-oscillator ground-state wavefunction. Importantly, these interaction coefficients are generically all of the same order in experiments, which leads to effectively infinite-range interactions along the synthetic dimension. This fundamental aspect precludes the physics of the conventional Bose-Hubbard model in laser-induced synthetic dimensions. While interactions can be tuned externally, using Feshbach resonances, the latter cannot directly produce the desired configuration, namely, strong and uniform intra-spin interactions $U_{mm}\!\approx\!U\!\gg\!U_{m\ne n}$ for all states $m$.

The aim of our scheme is to artificially generate on-site interactions along the synthetic dimension by pulsing strong intra-spin interactions $U_{mm}$ in a successive and time-periodic manner; see Fig.~\ref{fig1}(b). Specifically, one considers an experimental situation where one activates a strong interaction $U_{11}$ for a short duration $\tau$, then $U_{22}$ for the same duration $\tau$, etc \dots, while keeping the background interactions [see Eq.~\eqref{background} below] approximately constant and small. Physically, this requires enhancing the intra-spin scattering lengths, once at a time, in a periodic manner. As we demonstrate below, this protocol allows one to investigate the Bose-Hubbard model in synthetic dimensions, under the condition that the individual scattering lengths are modified on sufficiently small time scales (as compared to the physical time scales set by the effective Hubbard model).

This protocol can be formulated in terms of a Trotter sequence~\cite{Goldman-Dalibard}:~Considering one cycle of the driving protocol, the time-evolution operator over one period $T$ reads ($\hbar\!=\!1$ in the following except otherwise stated)
\begin{align}
&\hat{\mathcal{U}} (T)=\hat A_M(\tau) \,  \dots \, \hat A_m(\tau) \, \dots \, \hat A_2(\tau) \hat A_1(\tau) , \notag \\
&\hat A_m(\tau)=\exp(- i \tau \hat H_m ),
\label{trotter}
\end{align}
where $\tau\!=\!T/M$ is the duration of each pulse, and where the Hamiltonian at each step ``$m$" reads
\begin{equation}
\hat H_m=-\sum_{n=1}^M( \Omega_n \hat b^{\dagger}_{n+1} \hat b_{n}+\text{h.c.})+\frac{U}{2} \hat n_{m}(\hat n_{m}-1).\label{sequence}
\end{equation}
Hence, during the $m$th step, the system is assumed to evolve under the presence of the intra-spin interaction $U_{mm}\!=\!U$ only; we note that the scheme requires that the parameter $U$ should be independent of $m$. For the sake of simplicity, we have neglected the presence of the background (infinite-range) interactions, which at the $m$th step read
%\begin{align}
%&\hat H_m^{\text{bg}}=\sum_{n\ne m} \left ( \frac{U_{nn}}{2} \hat n_{n}(\hat n_{n}-1)+\sum_{l\neq n}\frac{U_{nl}}{2} \hat n_n \hat n_l \right ), \notag \\
%&\text{with} \quad U_{nn},U_{nl}\!\ll\!U_{mm}=U ,\label{background}
%\end{align}
\begin{align}
&\hat H_m^{\text{bg}}=\sum_{n\ne m} \frac{U_{nn}}{2} \hat n_{n}(\hat n_{n}-1)+\sum_{j,l\neq j}\frac{U_{jl}}{2} \hat n_j \hat n_l , \notag \\
&\text{with} \quad U_{nn},U_{jl}\!\ll\!U_{mm}=U ,\label{background}
\end{align}
the effects of which will be discussed in Section~\ref{sect_long_range}. Besides, we henceforth neglect the state-dependence of the coupling matrix elements, and set $\Omega_m\!=\!\Omega$.

When repeating the sequence in Eqs.~\eqref{trotter}-\eqref{sequence} in a fast and time-periodic manner, $\omega\!=\!2\pi/T\!\gg\!(\Omega,U)$, the (stroboscopic) time-evolution of the system is well captured by the effective (Floquet) Hamiltonian defined through~\cite{Goldman-Dalibard,Bukov}
\begin{equation}
e^{-i T \hat H_F}=\hat{\mathcal{U}} (T).
\label{effective_ham}
\end{equation}
Considering the standard Trotter expansion~\cite{Goldman-Dalibard}, one obtains a reasonable approximation for the effective Hamiltonian $\hat H_F$, 
\begin{equation}
\hat{\mathcal{U}} (T) \approx \exp(- i \tau \sum_m \hat H_m ) \equiv e^{-i T \hat H_{\text{TBH}}},\label{trotter_eq}
\end{equation}
which in this case yields the 1D Bose-Hubbard Hamiltonian
\begin{equation}
\hat H_{\text{TBH}}=-J_{\text{eff}} \sum_{m=1}^M(\hat b^{\dagger}_{m}\hat b_{m+1}+\hat b^{\dagger}_{m+1}\hat b_{m})+\frac{U_{\text{eff}}}{2}\sum_m \hat n_{m}(\hat n_{m}-1),
\label{eqf1}
\end{equation}
where we introduced the effective tunneling and interaction strengths 
\begin{equation}
J_{\text{eff}} = \Omega , \qquad  U_{\text{eff}}\!=\!U/M .
\end{equation} 
Summarizing, pulsing strong intra-spin interactions in a fast and time-periodic manner should allow one to investigate the physics of the Bose-Hubbard model in an atomic synthetic dimension, as captured by the effective ``Trotter-Bose-Hubbard" Hamiltonian $\hat H_{\text{TBH}}$ in Eq.~\eqref{eqf1}.

\section{Experimental implementation}\label{sect_exp}

The system discussed above can be implemented using a bosonic atom of hyperfine spin $F$, possessing $2F+1$ levels $\ket{m}$, with $-F\leq m\leq F$. The coupling $\Omega_m$ between successive levels $m$ and $m+1$ would be induced using a two-photon Raman transition. Independent tuning of the $\Omega_m$ couplings can be achieved by applying a quadratic Zeeman shift, which lifts the  degeneracy between the $m\leftrightarrow m+1$ transitions, allowing independent resonant couplings. Such a control over the coupling amplitudes can be used to vary the number of involved states $M\leq2F+1$.

In this work, we explore the strongly-interacting regime of the Trotter-Bose-Hubbard Hamiltonian in Eqs.~\eqref{trotter_eq}-\eqref{eqf1}, which corresponds to having a ratio $U_{\text{eff}}/J_{\text{eff}}\!\approx\!1-10$. Realistic experimental values for the model parameters are $J_{\text{eff}}\!=\!\Omega\!\approx\!2\pi\times 100$Hz and $U\!=\!U_{\text{eff}} M\!\approx\!2\pi\times1$kHz, which should allow for realistic observation times of a few hopping period, $t\!\approx\!1-10 (1/J_{\text{eff}})$, while limiting interaction-induced losses.

For the proposed scheme, the various intra-spin scattering lengths need to be dynamically tuned and modulated at a drive frequency $\omega\gg\!(J_{\text{eff}},U_{\text{eff}})$; see Section~\ref{sect_numerics}. Following the discussion above, this would correspond to having $\omega$ of the order of several kHz. This requires a fast control over the associated Fano-Feshbach resonances~\cite{chin_feshbach_2010}. This could be achieved by using standard magnetic tuning on the resonances, which would allow, if carefully implemented, for drive frequencies up to the order of the bandwidth of the magnetic field circuits (limited by the coils' inductance and eddy currents), typically a few tens of kHz~\cite{Clarck:2017,Clarck PHD Thesis}. In addition, the tuning of several scattering lengths via the magnetic field is intrinsically coupled, bringing additional correction to the proposed scheme (see also Section~\ref{sect_long_range}). Better adapted tuning schemes can also be envisioned, such as schemes based on optical beams, radiofrequency fields or microwave radiations~\cite{Fedichev:1996,Fatemi:2000, Theis:2004,Moerdijk:1995, Papoular:2010}. Such schemes are not only intrinsically fast but also allow for an independent and simultaneous tuning of several scattering lengths. The control of multiple Feshbach resonances in spin mixtures was first proposed in Ref.~\cite{zhang_independent_2009}, and was demonstrated experimentally with  $^{87}$Rb atoms in  \cite{kaufman_radio-frequency_2009}.

\section{Numerical investigations}\label{sect_numerics}

The next Sections analyze the validity and applicability of this scheme, in the relevant situation of a small ensemble of $N$ atoms. In order to explore the physics of the Bose-Hubbard model at unit filling ($\bar{n}\!=\!1$), we will henceforth set $N\!=\!M$, where $M$ corresponds to the number of available internal states, i.e.~the number of sites along the synthetic dimension. As reminded above, the number of involved internal states can be adjusted by controlling individual couplings, and it can thus be made even or odd.

The emblematic Bose-Hubbard model has played a key role in our exploration of quantum matter, by providing a minimal setting by which a genuine quantum phase transition can be observed:~the transition from a superfluid to a Mott insulating state~\cite{fisher,Sheshadri,Jaksch,greiner}. At unit filling, the Mott state is characterized by the occupation of each site by one particle; this feature is signaled by a small average value of the operator  
\begin{equation}
\hat n_D=\frac{1}{N}\sum_m\frac{\hat n_m(\hat n_m-1)}{2},
\label{nd}
\end{equation}
which counts the average number of doubly occupied sites. Besides, the Mott state is also associated with a non-local ``parity" parameter~\cite{Berg,endres}, which signals the appearance of correlated particle-hole pairs, as captured by the average value of the operator~\cite{endres}
\begin{equation}
\hat O_p(l)=e^{i \pi\sum_{m<l}(\hat n_m-\bar{n})},
\label{op}
\end{equation}
where $\bar{n}\!=\!1$ denotes the average background density, and where $l$ denotes a generic lattice site. In contrast, the superfluid phase is characterized by a vanishing gap (in the thermodynamic limit), a significant number of doubly occupied sites and a zero parity parameter $\langle \hat O_p(l) \rangle$. At unit filling ($\bar{n}\!=\!1$), the ground-state of the Bose-Hubbard Hamiltonian in Eq.~\eqref{eqf1} forms a Mott state when $U_{\text{eff}}\!\gtrsim\!3 J_{\text{eff}}$ and a superfluid otherwise; see Refs~\cite{kuhner1,kuhner2,kollath}.

\subsection{Accuracy of the Trotter-Bose-Hubbard Hamiltonian}\label{sect_accuracy}

In the present Trotter approach, we expect that accuracy is reached in the limit $\tau\!\rightarrow\!0$, namely, when the frequency of the periodic drive $\omega$ substantially exceeds all other frequencies in the system:~as $\hbar \omega$ becomes comparable to other energy scales (set by the model parameters $U_{\text{eff}}$ and $J_{\text{eff}}$), one expects corrections to the ``ideal" (Trotter) Hamiltonian in Eq.~\eqref{eqf1} to be significant~\cite{Goldman-Dalibard,heyl}. 

As a first step, we explore the accuracy of the Trotter-Bose-Hubbard Hamiltonian in Eq.~\eqref{eqf1}, by performing numerical (t-DMRG) simulations~\cite{white1,note2} of the dynamical evolution of a system of $N\!=\!M$ atoms under the full-time-dependent sequence in Eqs.~\eqref{trotter}-\eqref{sequence}. In these simulations, the initial state is taken to be the ground-state of the Trotter-Bose-Hubbard Hamiltonian in Eq.~\eqref{eqf1}, and the time-evolution operator associated with the multi-step sequence in Eqs.~\eqref{trotter}-\eqref{sequence} is applied in a repeated manner at the frequency $\omega$. We then compute the average values of both the local [Eq.~\eqref{nd}] and the non-local [Eq.~\eqref{op}] observables. In these calculations, we set $U_{\text{eff}}/J_{\text{eff}}\!=\!1$.

\begin{figure}
\includegraphics[scale=0.425]{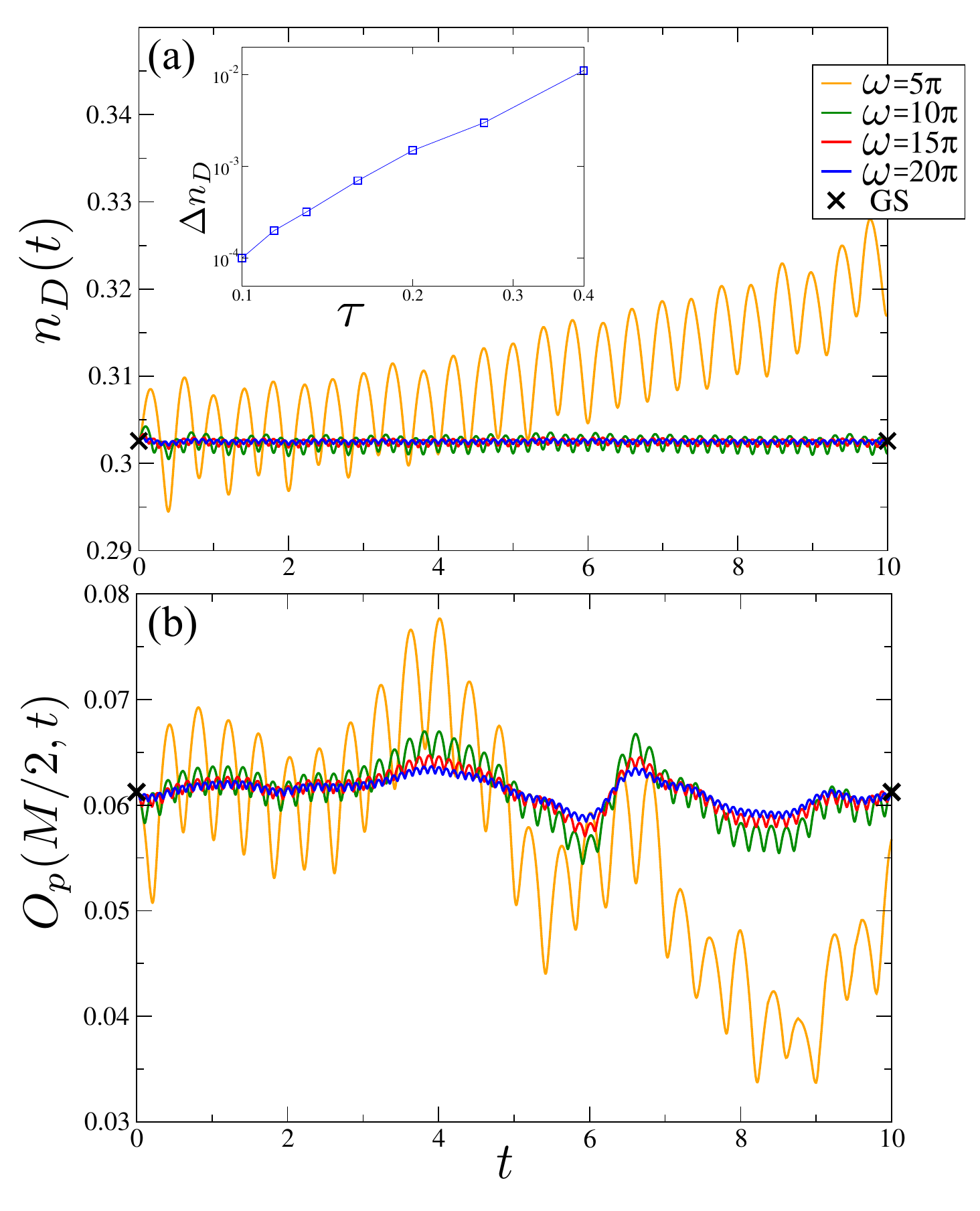}
\caption{Time-evolution of (a) $n_D(t)$ and (b) $O_{p}(M/2;t)$ for fixed $M=N=10$, $U_{\text{eff}}/J_{\text{eff}}=1.0$ and different values of $\omega$ (in units of $J_{\text{eff}}$). The black X indicate the ground state values for the same quantities, as obtained by diagonalizing Eq.~\eqref{eqf1}. The inset shows the quadratic growing of $\Delta n_D$ for different periods. The time $t$ is expressed in units of $1/J_{\text{eff}}$ and the frequency $\omega$ in units of $J_{\text{eff}}$.}
\label{fig2}
\end{figure}

As shown in Fig.~\ref{fig2}, for a sufficiently large drive frequency (here $\omega\!\gtrsim\!10\pi J_{\text{eff}}$), both observables undergo a small micro-motion around the initial (ideal) value associated with the ground-state of the effective Hamiltonian in Eq.~\eqref{eqf1}, i.e.~$n_D^{\text{ideal}}\!=\!n_D(t\!=\!0)$. For a smaller frequency $\omega\!\lesssim\!5 \pi J_{\text{eff}}$, one observes substantial deviations from the ideal values in the form of a larger micro-motion, as well as a dynamical increase in the mean double occupancy.  

We further study the accuracy of our method by plotting the relative error $\Delta n_D\!=\!|n_D(t)-n_D^{\text{ideal}}|$ as a function of the Trotter time-step $\tau$, which shows a characteristic quadratic growth~\cite{heyl}; see the inset in Fig.~\ref{fig2}(a). A similar behavior was found for the non-local observable in Eq.~\eqref{op}. 

Altogether, these studies offer a first confirmation that the Trotter sequence in Eqs.~\eqref{trotter}-\eqref{sequence} indeed effectively reproduces the time-dynamics associated with the Bose-Hubbard Hamiltonian, offering a viable scheme by which the latter model could be explored in an atomic synthetic dimension.

\subsection{Effects of the residual infinite-range interaction}\label{sect_long_range}

As previously noticed, the effective Bose-Hubbard Hamiltonian in Eq.~\eqref{eqf1} was derived upon making the assumption that the background ``infinite" range repulsion in Eq.~\eqref{background} can be neglected. In this paragraph, we study the range of validity of such an approximation, by taking a finite background interaction into account during each step of the Trotter sequence. For the sake of simplicity, we take this background interaction to be constant and set $U_{n n}\!=\!U_{j l}\!=\!V$ in Eq.~\eqref{background}. 
\begin{figure}
\includegraphics[scale=0.3]{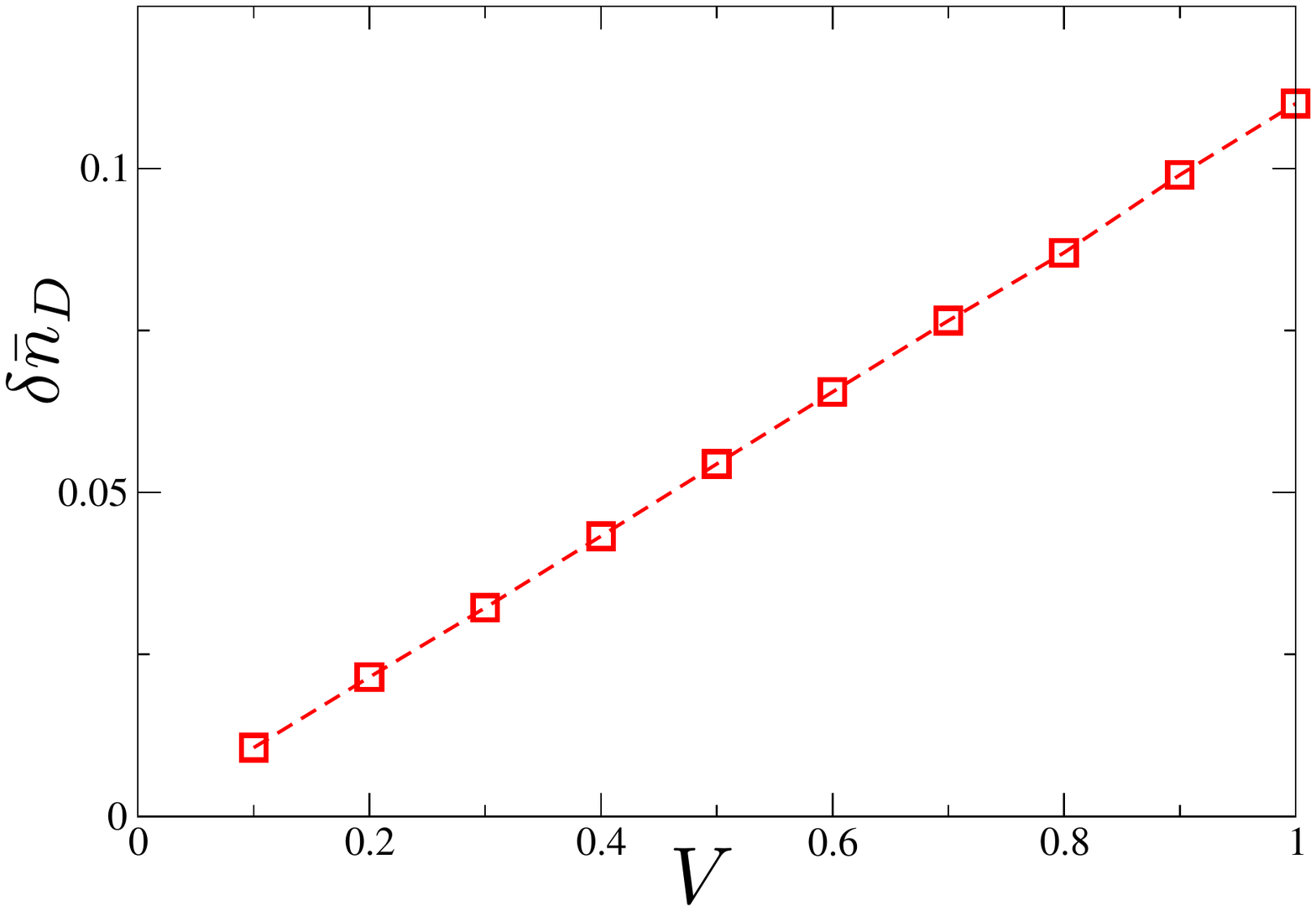}
\caption{$\delta \bar{n}_D$ for different values of the residual background interaction strength $V$ (in units of $J_{\text{eff}}$), as obtained by fixing $M=N=10$, $\omega=20\pi J_{\text{eff}}$ and $U_{\text{eff}}/J_{\text{eff}}=1.0$.}
\label{fig3}
\end{figure}

To analyze the effects of the residual background interaction, we compute the following indicator
\begin{equation}
\delta \bar{n_D} (V)=\frac{|\bar{n}_D(V=0)-\bar{n}_D(V)|}{\bar{n}_D(V=0)+\bar{n}_D(V)},
\label{err}
\end{equation} 
where $\bar{n}_D$ denotes the time average of the mean double occupancy number [Eq.~\eqref{nd}]. The numerical results shown in Fig.~\ref{fig3} indicate that a significant discrepancy ($\sim10\%$ error) with respect to the ideal case ($V\!=\!0$) only occurs for a strong residual background interaction $V\!\gtrsim\!U_{\text{eff}}$. We note that a similar behavior was found when analyzing the non-local observable in Eq.~\eqref{op}. Altogether, these results indicate that the residual interaction indeed plays a negligible role when pulsing strong intra-spin interactions with $U_{mm}\!\gg\! V \times M$. In particular, this suggests that our scheme is indeed well suited to explore the strongly-interacting regime of the Bose-Hubbard model where the effective interaction strength $U_{\text{eff}}$ sets the largest energy scale, and where the residual interaction $V$ can thus be safely neglected. In the following, we therefore set $V\!=\!0$.

%%%%%%%%%%%%%%%%%%%%%%%%%%

\subsection{From the superfluid to the Mott state \\ in an atomic synthetic dimension}

\begin{figure}
\includegraphics[scale=0.65]{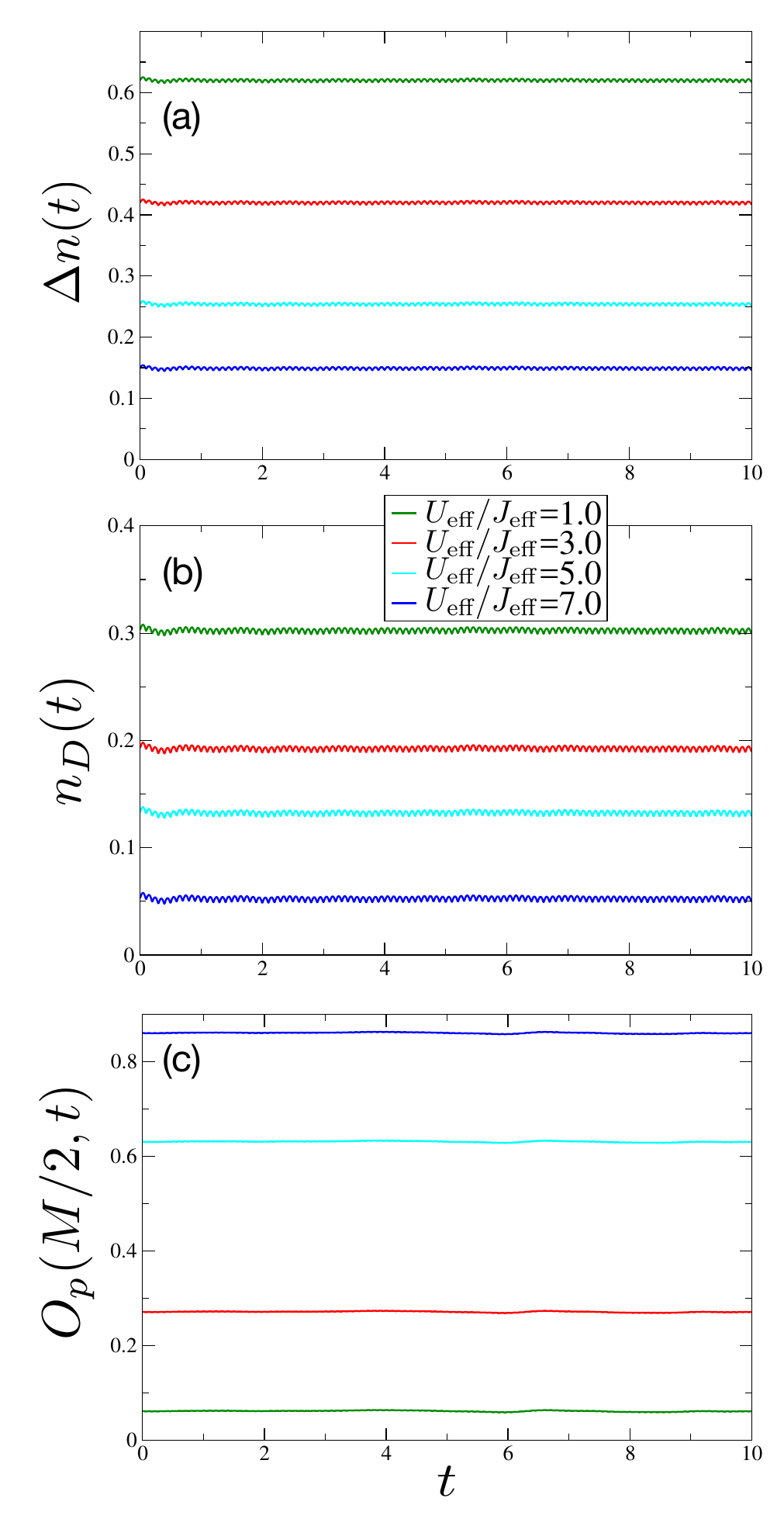}
\caption{ Time evolution of (a)  $\Delta n(t)$, (b) $n_D(t)$, (c) $O_p(M/2,t)$ for fixed $M=N=10$, $\omega=20\pi J_{\text{eff}}$ and different values of $U_{\text{eff}}/J_{\text{eff}}$. The time $t$ is expressed in units of $1/J_{\text{eff}}$.}
\label{fig4}
\end{figure}

The accuracy between the real time evolution associated with the Trotter sequence [Eqs.~\eqref{trotter}-\eqref{sequence}] and that associated with the target (ideal) Bose-Hubbard Hamiltonian in Eq.~\eqref{eqf1} was previously confirmed in the superfluid regime corresponding to $U_{\text{eff}}\!=\!J_{\text{eff}}$. We now validate this agreement through the superfluid-to-Mott transition, upon increasing the effective interaction strength. In our numerics, we evaluate the local and non-local observables that were previously introduced in Eqs.~\eqref{nd}-\eqref{op}, as well as the compressibility defined as
\begin{equation} 
\Delta n(t)=\frac{1}{M}\sum_{m}(\langle \hat n_m^2\rangle-\langle \hat n_m\rangle^2),
\label{com}
\end{equation}
which provides a further probe to distinguish between the two competing phases:~in the thermodynamic limit, the superfluid phase is gapless (compressible), while the Mott state is gapped (incompressible). Figure~\ref{fig4} shows the time-evolution of these three observables, when preparing the initial state in the ground-state of the effective Hamiltonian and acting on this state with the Trotter sequence. Importantly, we observe that all three observables undergo a very small micro-motion around their ideal (initial) value, for all values of the ratio $U_{\text{eff}}/J_{\text{eff}}$ considered. While the transition from the superfluid to the Mott state does not show a sharp behavior, which is indeed expected in such a small system size ($M\!=\!10$), one does recognize the main features of this quantum phase transition:~indeed, both the double-occupancy $n_D(t)$ and the compressibility $\Delta n(t)$ show smaller values upon crossing the expected transition point [$U_{\text{eff}}/J_{\text{eff}}\!\gtrsim\!3$], while the parity parameter $O_p(t)$ shows an opposite behavior (i.e.~significantly larger values are obtained in the Mott regime). We point out that the sharpness of this superfluid-to-Mott transition can be improved by considering more internal states ($M\!>\!10$), i.e.~a longer synthetic dimension.

\subsection{The flux ladder configuration: \\ Combining a synthetic dimension with a double well}

In this Section, we explore the possibility of extending the dimensionality of the effective Bose-Hubbard model in Eq.~\eqref{eqf1}, by considering atoms trapped in a double-well potential. Upon taking the synthetic dimension into account, this configuration leads to a fictitious ladder geometry that is aligned along the synthetic dimension:~hopping from a well to the other corresponds to hopping on the rungs of the ladder, while ``hopping" along the internal-state (synthetic) dimension corresponds to moving along the legs of the ladder. As already pointed out in Ref.~\cite{celi}, the hopping matrix elements along the synthetic dimension are naturally associated with complex phase factors (which are attributed to the phases of the Raman lasers), which in turn generates synthetic magnetic fluxes in each plaquette of the ladder. 

Applying the Trotter sequence in Eqs.~\eqref{trotter}-\eqref{sequence} to this double-well setting leads to the effective Hamiltonian
\begin{eqnarray}
\hat H=&-&J_{\text{eff}}\sum_{m,\sigma}(e^{i\theta\sigma} \hat b_{m,\sigma}^{\dagger} \hat b_{m+1,\sigma}+\text{h.c.})
\nonumber\\
&-&\frac{J_{\text{real}}}{2}\sum_{m}(\hat b^{\dagger}_{m, \frac{1}{2}} \hat b_{m, - \frac{1}{2}}+ \text{h.c.})
\nonumber\\
&+&\frac{U_{\text{eff}}}{2}\sum_{m,\sigma}\hat n_{m,\sigma}(\hat n_{m,\sigma}-1),
\label{ladder}
\end{eqnarray}
where we have again neglected the residual background infinite-range interaction in Eq.~\eqref{background}. Here, in addition to the parameters $J_{\text{eff}}$ and $U_{\text{eff}}$ already defined in Eq.~\eqref{eqf1}, the Hamiltonian includes the parameter $J_{\text{real}}$ that describes tunneling between the two wells (which we label as $\sigma\!=\!\pm1/2$). The hopping matrix elements along the synthetic dimension also contain a phase factor $\exp (\pm i\theta /2)$, which explicitly depends on the well (or the leg of the ladder) $\sigma$. We note that a similar synthetic ladder configuration was realized using two long-lived electronic states of fermionic Ytterbium atoms~\cite{livi}, in which case the long (leg) direction was associated with a real (optical) lattice. 

The bosonic ladder Hamiltonian in Eq.~\eqref{ladder} has a rich ground-state phase diagram, featuring various types of so-called Meissner and vortex states~\cite{Atala,greschner1,greschner2,Piraud,didio,orignac}, which can be identified by evaluating the currents associated with the leg (synthetic) and rung (real) directions. In the following, we study the existence of such phases in the case of a particle density $\bar{n}\!=\!N/(2M)\!=\!1/2$ in the synthetic ladder. Following Ref.~\cite{greschner2}, we analyze the ``chiral" current $\hat {\cal{J}}_c$ and the local ``rung" current $\hat {\cal{J}}_r$ defined as
\begin{align}
&\hat {\cal{J}}_c=-iJ_{\text{eff}}\sum_{m,\sigma}\big(\sigma e^{i\theta\sigma}\hat b_{m,\sigma}^\dagger \hat b_{m+1,\sigma}-\text{h.c.}\big),
\label{jr} \\
&\hat {\cal{J}}_r=- \frac{i J_{\text{real}}}{M}\sum_{m\in \text{center}}\big(\hat b_{m,\frac{1}{2}}^\dagger \hat b_{m,-\frac{1}{2}}-\text{h.c.}\big),
\label{js}
\end{align} 
where ``$m\in \text{center}$" refers to a few selected ``sites" around the central ``site" $m\!=\!M/2$. We point out that the chiral-current operator associates different flow directions with the two legs of the ladder ($\sigma$), hence probing the orientation (or chirality) of the flow taking place along the edge of the ladder. In contrast, the local rung current operator probes vortices in the ladder, which are associated with flows taking place around a few inner plaquettes; see Refs.~\cite{greschner1,greschner2,Piraud,didio,orignac}.

In the Mott regime, the Meissner-type phase is characterized by a large ``chiral" current and by a vanishing rung current:~this phase is associated with a clear chiral motion along the legs of the ladder. In contrast, the Mott-vortex phase displays a significant rung current accompanied with a reversal of the chiral current; see Fig. 5 in Ref.~\cite{greschner2}. 

\begin{figure}
\includegraphics[scale=0.3]{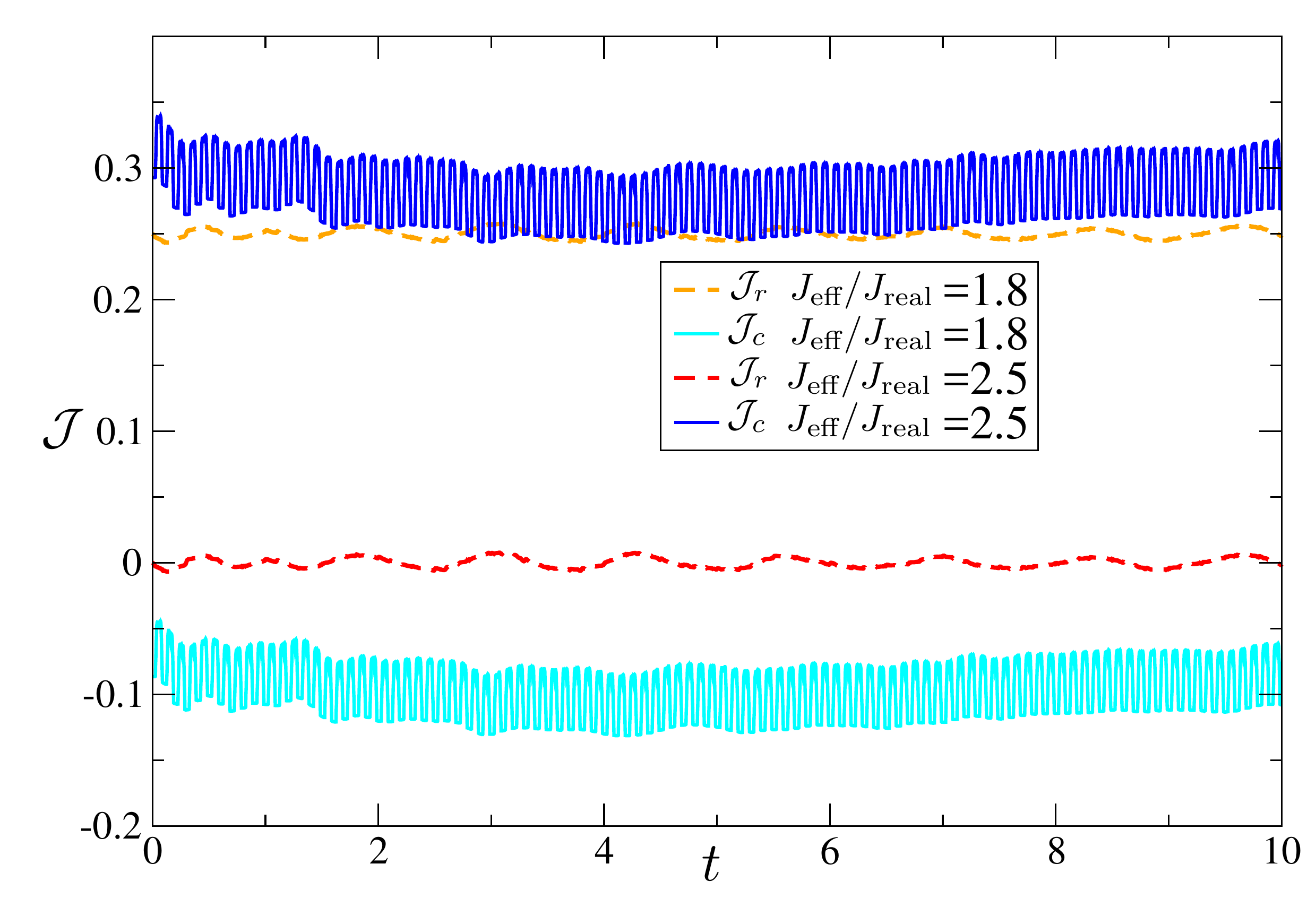}
\caption{Time evolution of the rung current ${\cal{J}}_r$ and chiral current ${\cal{J}}_c$ for a synthetic ladder with a density $N/2M=0.5$ with $M=N=8$, $\omega=20\pi J_{\text{eff}}$, $\theta=0.9\pi$ and $U_{\text{eff}}/J_{\text{eff}}=4$ and different values of $J_{\text{real}}/J_{\text{eff}}$. The local rung current is evaluated using the four central sites. The time $t$ is expressed in unit of $1/J_{\text{eff}}$ and the chosen initial state corresponds to the ground state of Eq.~\eqref{ladder}.}
\label{fig5}
\end{figure}

We demonstrate in Fig. \ref{fig5} that such phases are indeed supported by the interacting synthetic ladder generated by the Trotter sequence in Eqs.~\eqref{trotter}-\eqref{sequence}. This figure shows the time-evolution of the chiral and rung currents, when preparing the initial state in the ground-state of the effective Hamiltonian in Eq.~\eqref{ladder} and acting on this state with the Trotter sequence. As in the previous examples, one observes a very weak micro-motion of these observables around their ideal (initial) value, which coincide with those previously reported in Ref.~\cite{greschner2}:~Depending on the value of $J_{\text{real}}/J_{\text{eff}}$, the states have large (resp.~weak) mean chiral (resp. rung) current or a large rung current accompanied with a weak and negative chiral current.  

These results clearly show that both types of phases (Meissner and vortex) are indeed supported by the Trotter-engineered interacting ladder, in the Mott regime ($U_{\text{eff}}/J_{\text{eff}}\!>\!3$). An interesting perspective concerns the realization of Laughlin-type states (fractional Chern insulators) in this setting~\cite{FCI}, by extending the size of the ``real" direction.

\subsection{Quench dynamics and magnetic order \\ in a synthetic dimension}

%%%%%%%%%%%%

The previous Sections explored the ground-state properties of the effective Trotter-Bose-Hubbard Hamiltonian [Eq.~\eqref{eqf1}]. Specifically, those Sections analyzed the time-evolution of a specific initial state, taken to be the ground-state of the effective Hamiltonian, upon applying the Trotter sequence in Eqs.~\eqref{trotter}-\eqref{sequence}. In this last Section, we now demonstrate that this interacting synthetic-dimension setting is also well suited to explore intriguing quench dynamics, namely, situations where the initial state does not correspond to an eigenstate of the effective Hamiltonian. 

Of particular interest are quench dynamics involving dynamically-formed bound states -- so-called bound pairs (BPs) in the case of contact interactions -- which can be realized in ultracold atoms through a special state preparation. Let us first recall the basic principles behind this concept: Consider a pair of atoms initially occupying a single lattice site; the energy difference between this configuration and that corresponding to spatially-separated atoms is then simply given by the onsite (Hubbard) interaction. When the latter is strong enough (irrespective of its sign), energy conservation forces the pair to remain bound:~the pair is forbidden to decay by converting its potential energy into kinetic energy due to the fact that the latter is restricted by the bandwidth of the populated Bloch band. In this regime, the prepared pair moves over the lattice, with both atoms tunneling together to the same neighboring sites. Such scenarios were theoretically proposed to create effective antiferromagnetic states~\cite{muth}, entangled Bell pairs \cite{keilmann} as well as metastable fermionic superfluids~\cite{rosch}. On the experimental side, such quench dynamics allowed for the observation of
repulsive BPs~\cite{winkler,strohmaier}, and magnon dynamics~\cite{fukuhara}. We also note that similar scenarios can arise in the regime of attractive interactions~\cite{mark}.

It is the aim of this Section to analyze the quench dynamics of bound pairs in atomic synthetic dimensions with engineered on-site interactions [Eqs.~\eqref{trotter}-\eqref{sequence}]. To this purpose, we consider an initial state of the form
\begin{equation}
\vert \psi (t=0) \rangle= | n_{1},n_{2},...n_{M}\rangle=|0,2,0,2,....0,2\rangle, 
\label{state}
\end{equation}
where $n_{m}$ corresponds to the (initial) number of atoms in the internal state $m$. This initial configuration corresponds to having a (synthetic) lattice with alternating doubly-occupied/empty lattice ``sites" (with no single-occupancy). We consider acting on this initial state with the Trotter sequence in Eqs.~\eqref{trotter}-\eqref{sequence}, which was previously shown to mimic the time-evolution associated with the effective Bose-Hubbard Hamiltonian in Eq.~\eqref{eqf1}. In this Section, we also set the interactions to be \emph{attractive}~\cite{mark}, i.e.~$U_{\text{eff}}\!<\!0$. In this configuration, we expect the initial pairs (in internal states with $m$ even) to remain bound in the synthetic dimension for sufficiently strong interactions $\vert U_{\text{eff}}/J_{\text{eff}}\vert\!\gg\!1$, and to ``tunnel" to neighboring ``sites" $m$ with an effective hopping matrix element smaller than the single-particle hopping matrix element $J_{\text{eff}}$. 

In this configuration where attractive bound pairs are formed and no single occupation is present, the system is captured by an effective model~\cite{muth}
\begin{equation}
\hat H=-J_{\text{BP}}\sum_{m}(\hat c_m^\dagger \hat c_{m+1}+\text{h.c.})+B\sum_m \hat c_m^\dagger \hat c_{m+1}^\dagger \hat c_{m+1} \hat c_{m},
\label{rbpham}
\end{equation} 
where $\hat c^\dagger_{m}$ creates a pair of atoms (a hard-core boson) in the $m$th internal state, where $J_{\text{BP}}\!=\!-2J_{\text{eff}}^2/U_{\text{eff}}$ is the ``nearest-neighbor" pair hopping, and where $B\!=\!-16J_{\text{eff}}^2/U_{\text{eff}}$ describes the density-density interaction strength between two neighboring pairs. Interestingly, a simple transformation $\hat S^z_m\!=\!(1-\hat n_m)/2$ maps this Hamiltonian onto a spin chain reminiscent of the XXZ model. In particular, for $B/J_{\text{BP}}\!>\!1$, the systems is known to undergo a phase transition from a paramagnetic to an antiferromagnetic state; in terms of the original bosons, this corresponds to a transition from a superfluid to a charge density wave. 

\begin{figure}
\includegraphics[scale=0.35]{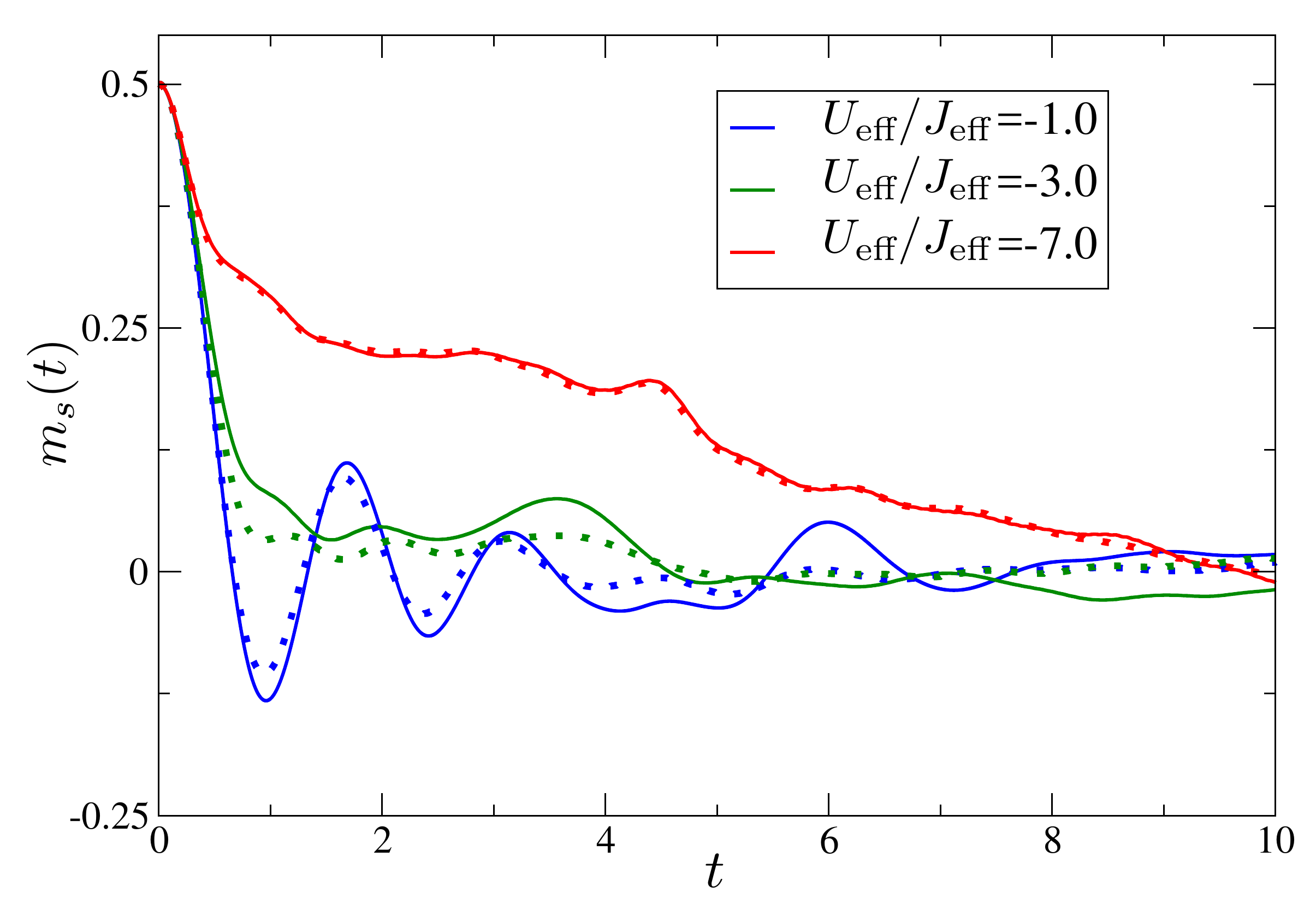}
\caption{Time-evolution of the staggered magnetization $m_s$ for different values of $U_{\text{eff}}/J_{\text{eff}}$. In all the results we fixed $M=N=8$, $\omega=20\pi J_{\text{eff}}$ and the time $t$ is expressed in unit of $1/J_{\text{eff}}$. The chosen initial state corresponds $|n_{1},n_{2},...n_{M}\rangle=|0,2,0,2,....0,2\rangle$; see Eq.~\eqref{state}. Full lines show the dynamics associated with the Trotter sequence, while dotted lines show that generated by the effective Bose-Hubbard Hamiltonian in Eq.~\eqref{eqf1}, using the same initial state.}
\label{fig6}
\end{figure}

We now demonstrate that our interacting synthetic dimension supports dynamically-induced antiferromagnetic states of bound pairs. To do so, we consider the initial state in Eq.~\eqref{state} and monitor the time-evolution of the mean staggered magnetization \begin{equation}
\hat m_s=\frac{1}{N}\sum_m(-1)^m(\hat n_m-1),
\end{equation}     
under the Trotter sequence [Eqs.~\eqref{trotter}-\eqref{sequence}]. In the spin-1/2 representation, the quantity $\langle \hat m_s \rangle$ acts as an order parameter for the antiferromagnetic state, while it detects a density wave state in terms of the original bosons. Following Ref.~\cite{barmettler}, we expect the staggered magnetization to display oscillations around the value $m_s\!=\!0$ for a small ratio $B/J_{\text{BP}}$, the system effectively behaving as free fermions in this paramagnetic regime. In contrast, for $B/J_{\text{BP}}>1$, one expects $m_s$ to decay exponentially to zero, as a signature of the underlying antiferromagnetic order~\cite{barmettler}. This behavior is precisely recovered in our numerical results displayed in Fig.~\ref{fig6}, which shows the time-evolution of the staggered magnetization for various values of the ratio $U_{\text{eff}}/J_{\text{eff}}$. For sufficiently strong intra-spin interactions, the system dynamically forms bound pairs in the synthetic dimension, hence leading to an underlying magnetic order and an approximately exponential decay of the staggered magnetization over time as captured by the effective Hamiltonian in Eq.~\eqref{rbpham}. We point out that a comparison between the time-evolution associated with the Trotter sequence (full lines in Fig.~\ref{fig6}) and the dynamics generated by the effective Bose-Hubbard Hamiltonian in Eq.~\eqref{eqf1} (dotted lines in Fig.~\ref{fig6}), as obtained from the same initial state [Eq.~\eqref{state}], show excellent agreement. We finally note that these results are compatible with those presented in Ref.~\cite{barmettler}. 

These results confirm the viability of our scheme in view of observing dynamically-formed bound states and artificial magnetism using atomic synthetic dimensions.

\section{Concluding remarks}

This work introduced a realistic scheme by which strong and uniform ``on-site" interactions can be generated and controlled in synthetic dimensions based on atomic internal states. This novel feature significantly broadens the applicability of synthetic dimensions in view of realizing and observing strongly-interacting states of matter in ultracold gases. 

In this proposal, we have illustrated and validated this scheme using several examples, including the study of unusual phases in strongly-interacting ladders with flux and signatures of antiferromagnetic order upon a quench. Our numerical simulations, based on t-DMRG, confirm that these appealing phases and dynamics could indeed be explored in small atomic synthetic dimensions, using available technologies. 

An interesting perspective concerns the extension of the 1D Bose-Hubbard model treated in this work to higher-dimensional settings, which could be realized by combining an atomic synthetic dimension with a real dimension (i.e.~an optical lattice) or another synthetic dimension spanned by a secondary internal degree of freedom. This possibility, together with the engineered interactions introduced in this work, would offer a new scenario by which the physics of 2D Hubbard models (including fractional Chern insulators) could be studied in ultracold atoms. Another important direction concerns the realization of the Fermi Hubbard model in synthetic dimensions, using a generalization of the present scheme.

%\begin{acknowledgments}
{\it Acknowledgments:} We thank J. Budich, J. Dalibard, F. Ferlaino, R. Lopes, T. Scaffidi, L. Tarruell and P. Zoller for useful discussions. Work in Brussels is supported by the FRS-FNRS and the ERC through the Starting Grant project TopoCold. L.~C. acknowledges the support of the DFG/FWF via the FOR grant FOR2247/PI2790. S.~N. acknowledges support from the European Union through the ERC grants UQUAM and TOPODY.
%\end{acknowledgments}

%
%


\begin{thebibliography}{99}
\bibitem{note_collisions} For Zeeman sublevels, see Section~\ref{sect_exp}, spin-changing collisions can be energetically suppressed in the presence of a quadratic Zeeman field.
\bibitem{note2} All the t-DMRG results are obtained by fixing the time step $\delta t=10^{-3} (1/J_{\text{eff}})$ and keeping up to 1500 dmrg states in the time evolution.
\end{thebibliography}
\end{document}